\documentclass[11pt,twoside]{article}

\usepackage{asp2006}
\usepackage{epsf}
\usepackage{psfig}
\usepackage{lscape}

\begin{document}
\title{The Properties of Dense Molecular Gas in the Milky Way and Galaxies}   
\author{Yancy L. Shirley, Jingwen Wu, R. Shane Bussmann, Al Wootten}   
\affil{Univ. of Arizona, CfA, Univ. of Arizona, NRAO}    

\begin{abstract} 
We review the evidence for a constant star formation rate per unit
mass in dense molecular gas in the Milky Way and the extragalactic
correlations of $L_{IR}$ with L$^{\prime}$ from observations of 
dense molecular gas.
We discuss the connection between the constant SFR/M interpretation
in dense gas and the global Schmidt-Kennicutt star formation law.
\end{abstract}


\section{SFR/M in the Milky Way} 
The connection between star formation studied in the Milky Way and
in galaxies is important for understanding the global evolution of
stars in the universe.  Lessons learned locally, where angular resolution
permits detailed studies of high-mass star forming regions, may be
applied to star formation regions that are unresolved in other galaxies.
Star formation occurs in dense molecular gas (Evans 1999);  therefore,
it is necessary to study the properties of dense molecular gas to
understand the conditions in the material actively involved in forming
protostars.  An important aspect of this study is 
the efficiency of star formation or how effectively dense gas is
converted into stars.  

One method for determining the star formation 
efficiency in Milky Way molecular
clouds is to calculate the ratio of the bolometric luminosity to
the mass of a star-forming clump within a giant molecular cloud
($L/M$).  
Within high-mass star forming regions in the Milky Way, submillimeter
single-dish telescopes can resolve individual cluster-forming
clumps ($\theta_{mb} < \approx 30^{\prime\prime}$).  Higher resolution
interferometric observations resolve the clumps into individual
star-forming cores (e.g. Brogan et al. 2007).  Only a few
high-mass clumps have been observed at high resolution. 
Since several systematic single-dish surveys of clumps have been
made (e.g., Plume et al. 1997; Zinchenko et al. 2000; Sridharan et al. 2002),  
we focus on the properties of the cluster-forming clumps.
If we assume a universal stellar Initial Mass
Function (IMF) for all star-forming regions within the Milky Way,
then the bolometric luminosity is directly proportion to the
star formation rate ($L_{bol} \propto SFR$).
The virial mass of a molecular clump is calculated from
\begin{equation}
M_{\rm{virial}} = \frac{5 R \Delta v^2}{G \ln 2} 
\frac{a_{\rm{density}}}{a_{shape}} \; ,
\end{equation}
where R is the size of the clump, $\Delta v$ is the FWHM linewidth,
$a_{density} = (1 - 2p/5)/(1 - p/3)$ (for $p < 2.5$)
is the correction factor for a power-law density
of the form $n \propto r^{-p}$, and $a_{shape}$
is the correction
factor for an ellipsoidal shape (Bertoldi \& McKee 1989).
The size and shape are determined from the FWHM intensity contour of the
clump (see Shirley et al. 2003).  An optically thin
linewidth must be determined from observations of isotopomers
(e.g. H$^{13}$CN, C$^{34}$S, H$^{13}$CO$^+$) since the linewidth of common 
dense gas tracers 
can be significantly broadened
due to large optical depths (Philips et al. 1979). 
Radiative transfer modeling of optically thin dust continuum emission 
indicates that the average large scale structure of dense molecular 
clumps follows a single power-law density $n \propto r^{-p}$ (Mueller et al. 2002;
Williams et al. 2005), for which the correction to $M_{vir}$ can be substantial
($a_{density} = 0.6$ for a $r^{-2}$ power-law).

\begin{figure}[!ht]
\plotone{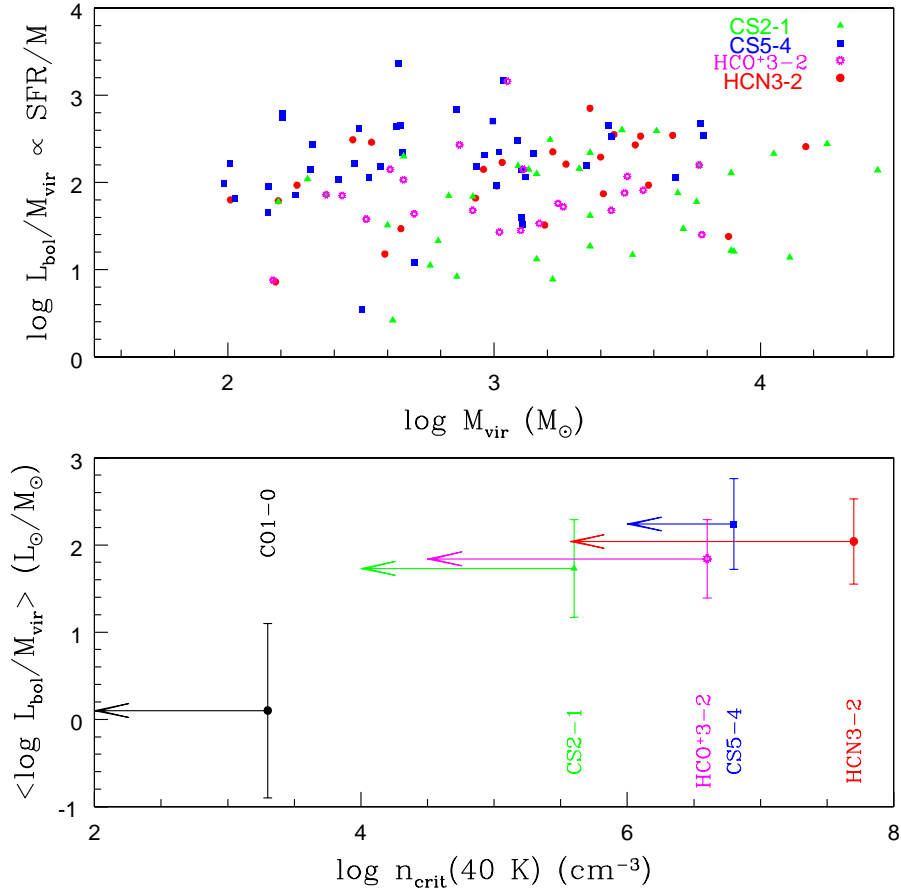}
\caption{$L_{bol}/M_{vir}$ is plotted for high-mass star forming clumps
observed in various dense molecular gas tracers (top panel).  The average
$\log L/M$ is plotted versus the critical density of the molecular tracer
calculated at $40$ K (bottom panel).  The arrows in density correspond
to the effective density for which a $1$ K line may be observed in
a typical clump (see Evans 1999).  Notice that $\langle \log L_{bol}/M_{vir} \rangle$ is
nearly constant for all molecular tracers above $10^4$ cm$^{-3}$.}
\end{figure}

The ratio of $L_{bol}/M_{vir}$ versus $M_{vir}$ is plotted
in Figure 1 for mapping surveys of CS J=2-1
and J=5-4 (Wu et al. 2008, in prep.; Plume et al. 1992; Shirley et al. 2003),
HCN J=3-2 (Wu \& Evans 2005), and HCO$^+$ J=3-2 (Shirley et al. 2008,
in prep.).  All of these dense gas tracers display a remarkable
lack of correlation.  When all of these tracers are considered
together, this result is consistent with a constant
SFR per unit mass of dense gas.  The average star forming efficiency 
traced by dense gas
is constant among high-mass star forming regions in the Milky Way and
this result appears to be independent of the dense gas tracer
observed (bottom panel, Figure 1).  
The critical density of a transition ($n_{crit} = A_{ul}/\gamma_{ul}$)
is crude approximation for determining the density at which a
particular transition is excited.  In reality, radiative transfer effects
can significantly lower the density
at which a particular line may be easily detected ($n_{eff} < n_{crit}$) by
up to an order of magnitude (see Evans 1999).  
The observed $\langle \log(L_{bol}/M_{vir}) \rangle$ is nearly constant for all 
tracers with
$n_{eff} > 10^4$ cm$^{-3}$. 
        
The result from molecular line observations is confirmed by making
$L_{bol}/M_{dust}$ comparisons from dust continuum observations.  Since the dust
emission on large scales is optically thin at (sub)millimeter
wavelengths, the dust mass may be calculated with assumptions of
the average dust temperature and the dust opacity
(Hildebrand 1983).  For instance,
the $350$$\mu$m radiative transfer modeling survey of Mueller et al. (2002)
find a constant $\log(L_{bol}/M_{dust})$ that is within a factor of
two of the $\log(L_{bol}/M_{vir})$ calculated from dense molecular gas 
(Shirley et al. 2003).
A factor of two is within the uncertainty in dust opacity
at long wavelengths in high-mass star forming regions.
Thus, there is strong evidence that, regardless of tracer,
the SFR/M is constant in dense molecular gas within the Milky Way.

\section{Dense Molecular Gas in Galaxies}

Observations of other galaxies are not currently able to resolve
individual cluster-forming molecular clumps;  since we cannot
calculate $M_{vir}$ directly, we must instead interpret global
averages of the molecular luminosity.  As a result, we calculate
$L^{\prime}$, or the source-integrated surface
brightness
\begin{equation}
L^{\prime} = 
3.256 \times 10^{-7} (D_L / \rm{Mpc})^2 (\nu_{rest} / \rm{GHz})^{-2}
(1+z)^{-1} \int S_{\nu} \, dv \rm{Jy \, km/s \, pc^2}
\end{equation}
where $D_L$ is the luminosity distance and $\nu_{rest}$ is the 
rest frequency
(see Mangum et al. 2007 for a derivation; Solomon et al. 1997).
The physical conditions in molecular gas may be probed by
comparison of $L^{\prime}$ for tracers with the same spatial extent.

Recent extragalactic surveys have found very strong correlations
between the globally integrated 
$L^{\prime}$ and the galactic infrared luminosity ($L_{IR}$, $\lambda \in [8,1000]$ 
$\mu$m).  The HCN 1-0 survey of Gao \& Solomon
(2004a,b) found a remarkable, tight
linear correlation between $L^{\prime}$(HCN1-0) and $L_{IR}$ 
over three orders of magnitude in $L_{IR}$.
This relationship is significantly different from the super-linear correlation
between $L^{\prime}$ and $L_{IR}$ observed for CO 1-0
(Figure 2).
Linear correlations are also observed in other dense gas tracers
including CO J=3-2 (Narayanan et al. 2005), 
HNC J=1-0, HCO$^+$ J=1-0, and CN N=1-0 and N=2-1 
(Baan et al. 2007; see Figure 2).  Furthermore, the 
extragalactic linear HCN $1-0$
correlation directly extends, without an offset, to $L^{\prime}$ versus $L_{IR}$ 
for galactic high-mass star-forming clumps (Wu et al. 2005).
If $L_{IR} \propto SFR$ and $L^{\prime} \propto M$, then
the linear correlation would indicate that the SFR in dense
molecular gas is also constant in other star-forming galaxies
($L_{IR}/L^{\prime} \approx$ constant, analogous with the galactic
$L_{bol}/M_{vir} \approx$ constant result).
One interpretation of these results is that the general
star formation law in other galaxies is a simple extension
of the constant SFR/M observed in the Milky Way with the dense
molecular clumps greater than a few hundred M$_{\odot}$ comprising
a fundamental unit of star formation (see Wu et al. 2005).
In this interpretation, the main difference between the Milky
Way and an extreme starburst galaxy is that the a larger
fraction of the molecular ISM is in a dense molecular phase in the
the starburst galaxy compared to the Milky Way.  Observationally, a larger 
ratio of HCN to CO 1-0 emission is observed with larger $L_{IR}$ 
(Gao \& Solomon 2004b; see Figure 2) possibly indicating a
higher dense gas fraction in more luminous galaxies (cf. Riechers et al.
2007 for a flattening of the ratio for high z galaxies).

There are several caveats to the constant SFR/M interpretation.  
L$^{\prime}$ is determined
globally and contains contributions from both low density
and high density molecular gas with a wide range of
excitation conditions along each line-of-sight.  $L^{\prime}$ may not be linearly 
proportional to mass; therefore, proper
interpretation of L$^{\prime}$ may required radiative transfer
modeling of excitation conditions in gas on galactic scales.
The observational study of Mangum et al. (2007)
attempts to circumvent this problem by calculated
the total molecular mass from LVG models of the  
absorption spectra of the centimeter K-doublet
H$_2$CO transitions.  This pilot survey detected
more than a dozen galaxies and was able to determine the
$L_{IR}/M_{H_2}$ ratio globally for an assumed formaldehyde abundance.  
The considerable scatter precludes strong conclusions, but the observed 
relationship may be consistent with a constant SFR/M (see Figure 2).

\begin{figure}[!ht]
\plotone{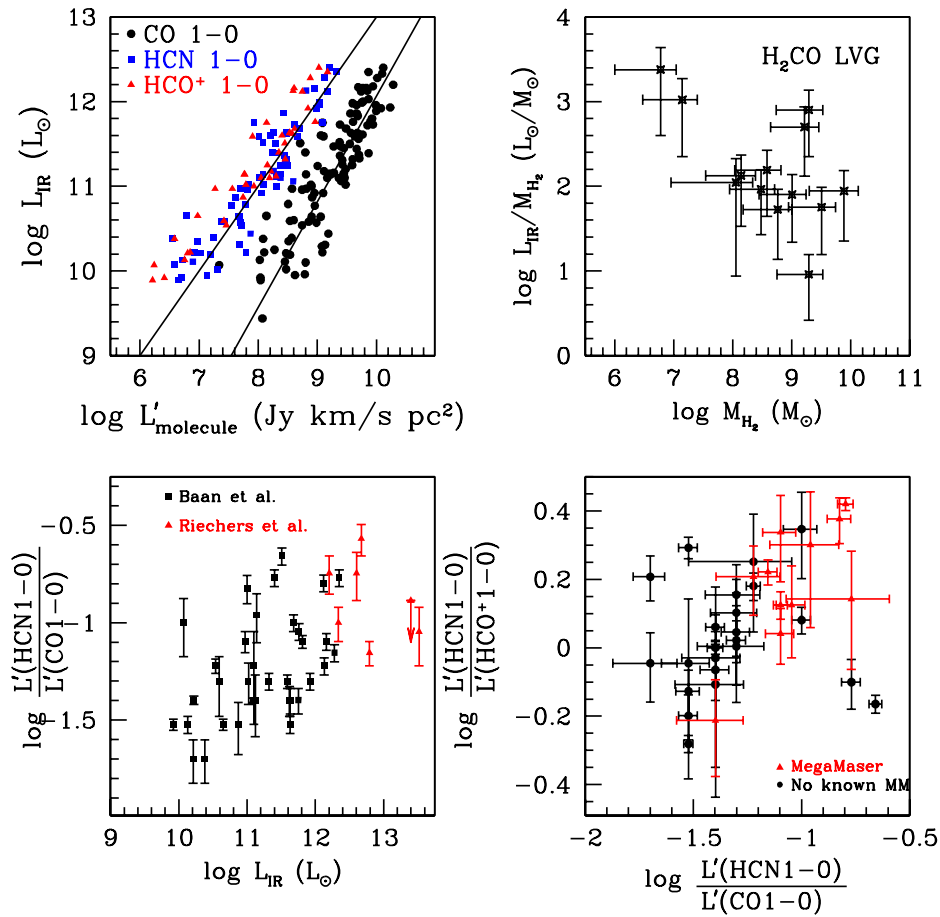}
\caption{Top left: $L_{IR}$ vs. $L^{\prime}$ for CO, HCN, and HCO$^+$ 1-0
($L^{\prime}$ calculated from 
data in Baan et al. 2007).  The solid lines have slopes of $1.0$ and $1.4$
to illustrate the difference between dense gas tracers (HCN and HCO$^+$)
and CO.  Top right: $L_{IR}/M_{H_2}$
vs. $M_{H_2}$ determined from LVG models of the lowest K-doublet transitions of
H$_2$CO observed by Mangum et al. 2007.  Bottom left: The ratio of HCN to CO
1-0 emission vs. $L_{IR}$ for galaxies compiled in Baan et al. (2007)
and Riechers et al. (2007; high-z detections).  Bottom right: The ratio of
HCN to HCO$+$ 1-0 emission vs. HCN to CO 1-0 emission (data compiled in
Baan et al. 2007).  Source with known megamasers (OH or H$_2$O), usually indicative
of AGN activity, are at the higher end of the correlation. }
\end{figure}

Theoretical calculations of $L^{\prime}$ must couple galactic hydrodynamics
with radiative transfer.  Two initial studies have provided
an alternative explanation to the observed linear correlations
for $L^{\prime}$(HCN1-0) and $L_{IR}$ and the super-linear correlation
of  $L^{\prime}$(CO1-0) and $L_{IR}$.  Krumholz \& Thompson (2007)
model the radiative transfer for an ensemble of clumps with
a lognormal distribution in density, as expected from
turbulent ISM simulations, and a SFR $\propto n^{1.5}$ based upon
free-fall arguments.  They find that the slope of the $L_{IR}$
versus $L^{\prime}$ correlation depends on whether the mean
density of the gas is above or below the critical density of
the tracer: $L_{IR}$ correlates super-linearly with $L^{\prime}$
if $\langle n \rangle > n_{crit}$ while $L_{IR}$ correlates linearly with $L^{\prime}$
if $\langle n \rangle < n_{crit}$.  Similar results were found
with the more sophisticated 3D coupled hydrodynamic-radiative transfer models
of Narayanan et al. (2007) with the main difference being
that Narayanan et al. also predict sub-linear correlations
for the higher $J$ transitions of dense molecular gas tracers (e.g. HCN 3-2).
This theoretical prediction is testable by current extragalactic surveys of 
higher excitation lines of HCN and HCO$^+$ (e.g., Padelis et al. 2007;
HHT survey of Bussmann et al. 2008, in prep.).
In both sets of models, $L^{\prime}$(HCN 1-0) is significantly affected
by sub-thermal excitation of 
large quantities of low density gas along lines-of-sight
through the galaxy.  Thus, the naive assumption
that $L^{\prime}$ faithfully traces mass may be incorrect.
Furthermore, both of these model assume an underlying star formation
law that is similar to the Kennicutt-Schmidt law of $SFR \propto
n^{1.5}$ (Kennicutt 1998; Kennicutt et al. 2007) and not
a constant $SFR/M$.

$L^{\prime}$ may also be affected by
the chemical or excitation effects on the dense molecular
gas via processes not related to star
formation such as the effects from an AGN (see Combes 2007 for
a chemical review).  AGN generate a substantial infrared
radiation field that may contribute to $L_{IR}$ (violating
the $L_{IR} \propto$ SFR assumption) and also may
enhance (sub)millimeter molecular transitions through absorption of
IR photons in mid-infrared vibrational bands (e.g., the $21$$\mu$m
bending mode of HNC; Aalto et al. 2007).
Active AGNs generate harder X-ray 
radiation fields (X-ray Dominated Regions, XDRs) 
than star forming regions and may effect global 
ionization balance and the chemistry in dense molecular gas.
For instance, a purported chemical trend has been observed in the 
ratio of HCN to HCO$^+$ emission versus the ratio of
HCN to CO emission that may be due to the effects of AGN
(see Figure 2; Graci\'a-Carpio et al. 2007;
Bann et al. 2007).  Comparison of the properties of the
galaxies with these higher molecular ratios tend to have evidence for strong
AGN activity (either via optical, x-ray, or megamaser identification). 
In one case, the Seyfert 2 galaxy NGC 1068 was imaged with
an interferometer and an enhancement in the HCN to CO ratio
is seen toward the central molecular torus (Krips et al. 2007).
Great care must be taken in the interpretation of these results
since theoretical prediction of the abundances in XDRs
is inchoate (e.g. Lintott et al. 2006; cf. Meijerink
et al. 2007). Multi-transition interferometric studies
with spatially resolved SED modeling are needed to fully
understand the excitation and abundance effects of the AGN on 
$L_{IR}$ and $L^{\prime}$.

\section{Summary}

Galactic observations indicate that the SFR per unit mass
traced by dense molecular transitions is constant,
independent of tracer above $n_{eff} > 10^4$ cm$^{-3}$.
In contrast, extragalactic observations of HI, CO, and
optical and near-infrared emission lines indicate that
the SFR follows a super-linear Schmidt-Kennicutt
law with $SFR \propto n^{1.5}$.  Recent radiative transfer
models of molecular emission on galactic scales indicate
that the observed linear correlations observed between
$L_{IR}$ and $L^{\prime}$ for dense gas tracers may
be related to an underlying Schmidt-Kennicutt law.   
Indeed, within the Milky Way, accounting of the gas content (HI + CO)
and cluster populations
within 1 kpc of the sun reveal that the Schmidt-Kennicutt
relationship predicts the observed star formation
surface density (Evans 2007).

How do we reconcile the two possibilities for the SFR law
(constant vs. super-linear or both)?  In the absence of direct 
observations of molecular clumps in 
other galaxies, the problem presents a difficult theoretical 
hydrodynamic-radiative transfer challenge which we have only begun 
to explore.  Within the Milky Way, it is still a theoretical
challenge to understand which process mitigate the observed
constant SFR/M in dense gas (see article by Mac Low in this volume).
On galactic scales, one intriguing possibility is that the 
local Schmidt-Kennicutt law may flatten at high densities such 
that the SFR $\propto n$.  This possibility should be explored theoretically.
Also, observation of higher excitation lines
at high spatial resolution with interferometers (e.g., PdBI, CARMA,
and ALMA) are needed to resolve issues of excitation and abundance 
variations within galaxies.  

The synthesis of current observational and
theoretical studies of dense molecular gas in 
the Milky Way and other galaxies is beginning to reveal 
the global properties of star formation in the gas that is actively involved 
in star formation.

\acknowledgements
YLS sincerely thanks Malcolm Walmsley, Daniela Calzetti, 
Mark Krumholz, Jeff Mangum, Desika Narayanan, and Stephanie Juneau for 
many stimulating discussions.


\end{document}